\documentclass[%
 reprint,
 nofootinbib,
 amsmath,
 amssymb,
 aps,
 prl,
 a4paper
]{revtex4-2}
\pdfoutput=1

\usepackage{graphicx}
\usepackage{dcolumn}
\usepackage{bm}
\usepackage{slashed}
\usepackage[dvipsnames]{xcolor}
\usepackage{physics}
\usepackage{comment}
\usepackage[normalem]{ulem}
\usepackage{hyperref}
\usepackage{url}
\usepackage{tikz}
\usepackage{mathrsfs}
\usetikzlibrary{quantikz2}

\begin{document}

\preprint{XXXX}
\title{Quantum Error Correction-like Noise Mitigation for Wave-like Dark Matter Searches with Quantum Sensors}

\author{Hajime Fukuda}
\email{hfukuda@hep-th.phys.s.u-tokyo.ac.jp}
\affiliation{Department of Physics, The University of Tokyo, Tokyo 113-0033, Japan}

\author{Takeo Moroi}
\email{moroi@hep-th.phys.s.u-tokyo.ac.jp}
\affiliation{Department of Physics, The University of Tokyo, Tokyo 113-0033, Japan}

\author{Thanaporn Sichanugrist}
\email{thanaporn@hep-th.phys.s.u-tokyo.ac.jp}
\affiliation{Department of Physics, The University of Tokyo, Tokyo 113-0033, Japan}

\date{\today}

\begin{abstract}
We propose a quantum error correction-like noise mitigation protocol for enhancing the sensitivity of wave-like dark matter searches with quantum sensors. Our protocol uses multiple sensors to mitigate the noise affecting each sensor individually, allowing for the suppression of excitation noise that is parallel to the dark matter signal. We demonstrate that our protocol can improve the sensitivity to dark matter signals by a factor of $\sqrt{N}$, where $N$ is the number of sensors used, {for small $N$}. Furthermore, {for sufficiently large $N$,} we find that our protocol achieves the same performance as the standard quantum limit by the ideal measurement, {which non-entangled sensors with parallel noise cannot reach} due to the unknown phase of the dark matter field.
Our work can be widely applied to various types of signals with unknown phases, and has the potential to enhance the sensitivity of quantum sensors such as arrays of resonant cavities.
\end{abstract}

\maketitle
\newpage

\textit{Introduction ---}
The dark matter (DM) in the universe is one of the most important mysteries in modern physics.
There are various candidates for DM, and one of the well-motivated candidates is light bosonic particles such as axions~\cite{Svrcek:2006yi,Arvanitaki:2009fg}, which is called ``wave-like DM'' due to its wave nature.
Recently, with the development of quantum technologies, growing attention has been paid to the use of quantum sensors for searching for wave-like DM\,\cite{Dixit:2020ymh,Chen:2022quj,Chen:2023swh,Ito:2023zhp,Chigusa:2023roq,Chen:2024aya,Braggio:2024xed,Chigusa:2024psk,Chigusa:2025rqs,Dong:2025mdk,Chen:2025tgj,Zhao:2025thg,Nakazono:2025tak,Bodas:2025vff,Banerjee:2025jss}.


{One of the important issues in using quantum sensors is the effect of noise parallel to the DM signal, which we call ``excitation noise''. With such parallel noise, it is known that we cannot outperform the standard quantum limit (SQL)\,\cite{huelga1997improvement,giovannetti2011advances,chaves2013noisy,kessler2014quantum,dur2014improved,arrad2014increasing,brask2015improved}. 
In the context of wave-like DM searches, even achieving the SQL is impossible. 
The ideal measurement, which saturates the quantum Cram\'er-Rao bound (QCRB)\,\cite{Paris:2008zgg}
and yields the SQL, reads out the DM-induced Rabi oscillation to first order in the interaction; e.g. for a two-level sensor with
a signal Hamiltonian proportional to the Pauli-$X$ operator, one can start from a Pauli-$Z$ eigenstate and perform a projective measurement in the Pauli-$Y$ eigenbasis, as we explain later.
However, the DM signal has an unknown phase that changes randomly over time scales longer than the coherence time of the DM field\,\cite{Foster:2017hbq,Fukuda:2025zcf}. 
Since the outcome of the ideal measurement depends on this varying phase (e.g., proportional to its sine or cosine), as we will explicitly see later, a simple average over long operations results in zero signal.
Thus, excitation noises significantly degrade the sensitivity of wave-like DM searches.
}

{
  The excitation noise at each sensor is independent while the DM signal is correlated.
  Therefore, it is natural to ask whether we can suppress the excitation noise by employing multiple sensors and analyzing their correlations.
}
Refs.\,\cite{Shu:2024nmc,Chen:2025tgj,Freiman:2025tse} have proposed to measure sensors with
entangled states such as $W$ states\,\cite{Dur:2000zz}, the equal superposition of single-excitation states across multiple sensors, to achieve this goal. Nevertheless, as discussed in Ref.\,\cite{Chen:2025tgj}, this procedure needs the presence of other noises, such as de-excitation noises {and cannot achieve the SQL in general}.
The underlying reason is that, when excitation noise is dominant, it can lead to multiple sensors being excited simultaneously --- events that the $W$-state measurement cannot distinguish. 

In this letter, we explore the possibility of processing sensor states during measurements to avoid this restriction. The crucial insight is that sensor states after noise effects occupy a subspace nearly orthogonal to the ``code space'', the space spanned by the $W$ state and the ground state. By designing a sensing protocol inspired by quantum error correction (QEC) techniques\,\cite{chaves2013noisy,kessler2014quantum,dur2014improved,arrad2014increasing,brask2015improved}, which pulls erroneous states back into the original code space repeatedly, we demonstrate that excitation noise can be mitigated even in the absence of other noise. {With $N$ sensors,}
the sensitivity to the DM signal is enhanced by a factor of $\sqrt{N}$ and the uncertainty scales as $1/N$ {for small $N$, and the sensitivity reaches the SQL by the ideal measurement, which has been considered unachievable in practical DM searches, for sufficiently large $N$.}
{In previous studies\,\cite{Sekatski:2017xdg,Demkowicz-Dobrzanski:2017stg,Zhou:2017kxr}, it has been considered that QEC-assisted metrology is useless when the noise and the signal Hamiltonian are parallel. However, we show that when we have unknown random parameters such as the phase of the DM field, QEC-like protocols can be useful even in such a case.}



\textit{DM Search with Quantum Sensors ---}
\label{sec:setup}
Let us first summarize our setup for DM searches with quantum sensors.
We consider $N$ quantum sensors, each of which is modeled as a two-level system with ground state $\ket{0}_i$ and excited state $\ket{1}_i$ for the $i$-th sensor; see App.\,\ref{app:notation} for definitions of states used in this letter. Note that multi-level systems such as resonant cavities can be treated in the same framework by focusing on two relevant energy levels\,\cite{Fukuda:2025zcf}.
The energy difference between the two states is denoted as $\omega_0$, so that the Hamiltonian of the $i$-th sensor is given by $H_i = -(\omega_0 / 2) \sigma_i^z$, where $\sigma_i^a$ ($a = x, y, z$) are the Pauli operators acting on the $i$-th sensor.
On the other hand, the DM field is modeled as a classical oscillating field with frequency $m$. The interaction between the DM field and the sensors is described by the Hamiltonian $H_{\rm int} = \epsilon \cos (m t + \varphi) \sum_{i=1}^N \sigma_i^x$, where $\epsilon$ is a small coupling constant and $\varphi$ is an unknown phase. DM field is assumed to be coherent over a time scale $\tau_\text{DM} \sim 2\pi / (m v^2)$, where $v \sim 10^{-3}$ is the typical velocity of DM particles in our galaxy, so that $\varphi$ can be treated as a constant during the time scale $\tau_\text{DM}$ but changes randomly for longer time scales. The same conclusion holds if we treat the DM field as a classical stochastic field\,\cite{Foster:2017hbq,Fukuda:2025zcf}.

To detect the DM signal, we start from the initial state $\ket{\psi_{\rm init}} = \bigotimes_{i=1}^N \ket{0}_i$ and let the system evolve under the total Hamiltonian $H = \sum_{i=1}^N H_i + H_{\rm int}$ for a time duration $\tau\lesssim 2\pi / (m v^2)$.
We move to the interaction picture with respect to $\sum_{i=1}^N H_i$ assuming that $|\omega_0 - m| \tau \ll 1$ so that we can apply the rotating wave approximation (RWA) and neglect the rapidly oscillating terms.
Then, the effective Hamiltonian in the interaction picture is given by
\begin{align}
    H_I = \epsilon \sum_{i=1}^N \qty( \sigma_i^x \cos \varphi -  \sigma_i^y \sin \varphi).
\end{align}
{Here, we assume the {displacement} of each sensor is much smaller than the de Broglie wavelength of the DM field so that the phase of the DM field is common to all sensors\,\cite{Fukuda:2025zcf}.}

In addition to the DM signal, we take into account the effect of environmental noise. For simplicity, we focus on the Markovian noises described by the Lindblad master equation\,\cite{Lindblad:1975ef,Gorini:1975nb}. In this letter,
{
  instead of the excitation noise, whose Lindblad operators are proportional to $\sigma_i^x + i \sigma_i^y$,
}
we consider the bit-flip noise in the interaction picture, whose Lindblad operators are $L_i = \sigma_i^x$ for $i=1,2,\ldots,N$.
{This does not lose generality as explained in App.\,\ref{app:reduction_to_bit_flip}; other types of noise can be reduced to the bit-flip noise by constructing logical qubits using the GHZ state\,\cite{Greenberger1989} from 3 physical qubits and applying QEC-like operations\,\cite{kessler2014quantum,dur2014improved,arrad2014increasing} to project noises to the bit-flip noise.}
The time evolution of the density matrix $\rho$ in the interaction picture is then given by
\begin{align}
    \frac{d\rho}{dt} &= \mathcal{L}[\rho] \equiv    -i [H_I, \rho] + \gamma \sum_{i=1}^N D_i[\rho], \label{eq:lindblad}\\
    D_i[\rho] &\equiv \frac{1}{2} \qty( \sigma_i^x \rho \sigma_i^x - \rho),
\end{align}
where $\gamma$ is the bit-flip rate and assumed to be the same for all sensors. 
{
  For DM signals that are on the verge of detection, the noise is stronger than the signal. Also, for the typical mass range of interest for wave-like DM searches, $1/m \lesssim 0.1\,\text{ns}$ and $\tau \lesssim 0.1\,\text{ms}$\,\cite{Chen:2022quj}. Thus,
}
we assume $\gamma \gg |\epsilon|$ and $\gamma \tau \lesssim 1$ in this letter.

\textit{QEC-like Error Mitigation Protocol ---}
\label{sec:protocol}
Throughout this letter, we assume that we may apply quantum operations such as unitary gates and measurements on the sensors in an arbitrary short time without introducing any error.
Under these assumptions, we propose a QEC-like protocol to mitigate the effect of the bit-flip noise. Our sensing protocol consists of sequentially repeating adaptive steps, such as Refs.\,\cite{Demkowicz-Dobrzanski:2014gvc,Sekatski:2017xdg,Demkowicz-Dobrzanski:2017stg,Zhou:2017kxr}.

\begin{figure}[tbp]
    \centering
  \begin{quantikz}
    \lstick{$\ket{0}_1$}
      & \gate{\mathcal{E}_{\Delta t}} \gategroup[3,steps=2,style={dashed, rounded corners, inner sep=2pt}]{Unit interval} & \gate[3]{\mathcal{E}_\text{c}} & \ \ldots\ &  \gate{\mathcal{E}_{\Delta t}} & \gate[3]{\mathcal{E}_\text{c}}  & \meter[3]{}\\
     \lstick{\ \vdots\ } \setwiretype{n} & \ \vdots\   & & \ \ldots\  & \ \vdots\  & &\\
      \lstick{$\ket{0}_N$}&  \gate{\mathcal{E}_{\Delta t}} & & \ \ldots\ & \gate{\mathcal{E}_{\Delta t}} & & 
  \end{quantikz}
    \caption{
      A schematic quantum circuit of our QEC-like error mitigation protocol.
    }
    \label{fig:protocol}
\end{figure}
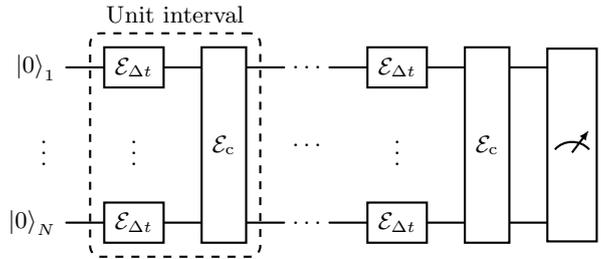

A schematic diagram of our protocol is shown in Fig.\,\ref{fig:protocol}.
We divide the sensing time $\tau$ into $N_\tau$ intervals of duration $\Delta t = \tau / N_\tau$.
$\Delta t$ is chosen such that $N \gamma \Delta t \ll 1$ and $m \Delta t \gg 1$,
so that the probability of having two or more bit-flip errors during each interval is negligible and the RWA is valid, respectively.
{In realistic situations, the energy relaxation time can be as long as $1/\gamma \sim 1\,\text{ms}$ for state-of-the-art quantum sensors such as superconducting qubits\,\cite{Bland:2025byl}. Thus, a wide range of parameters, e.g., $N \gtrsim 10^3$, can satisfy the above conditions. Future improvements in quantum sensor technology may allow for even larger $N$.}
Note that our protocol can be applied even if the RWA is not valid by performing a similar analysis as Ref.\,\cite{Sichanugrist:2024wfk}.
Each unit interval, shown as the dashed box in Fig.\,\ref{fig:protocol}, consists of two steps; first, we let the system evolve under the master equation for a time duration $\Delta t$, which is denoted as the quantum channel $\mathcal{E}_{\Delta t}$ in Fig.\,\ref{fig:protocol}. 
Since the time evolution is governed by the Lindblad equation, the quantum channel $\mathcal{E}_{\Delta t}^{\otimes N}$ is given by
\begin{align}
    \mathcal{E}_{\Delta t}^{\otimes N}[\rho(t)] = \rho(t + \Delta t) \simeq \rho(t) +  \mathcal{L}[\rho(t)] \Delta t.
    \label{eq:channel_E_dt}
\end{align}
Here, we have used the fact that $N \gamma \Delta t \ll 1$ to expand the time evolution up to the first order in $\Delta t$.
Next, we apply a control operation $\mathcal{E}_{\rm c}$ to mitigate the bit-flip errors that may have occurred during the time interval $\Delta t$. As we have claimed, we assume that this control operation takes negligible time and does not introduce additional noise. 

We now describe the construction of the control operation $\mathcal{E}_{\rm c}$ for mitigating bit-flip errors. The DM signal Hamiltonian $H_I$ excites $\ket{0} = \bigotimes_{i=1}^N \ket{0}_i$ to $\ket{W} = \frac{1}{\sqrt{N}} \sum_i \sigma_i^x \ket{0}$, motivating us to define the logical code space as $\mathcal{H}_L = \{\ket{0}_L \equiv \ket{0},\, \ket{1}_L \equiv \ket{W}\}$. We design $\mathcal{E}_{\rm c}$ to (approximately) detect and correct single bit-flip errors such that the state after $\mathcal{E}_{\rm c}$ lies within the code space.
Then, what can be the syndrome measurement and the recovery operation?
A bit-flip error on the $i$-th sensor transforms a code state $a\ket{0} + b\ket{W}$ into $a\ket{i} + b\ket{2,i} + \mathcal{O}(N^{-1/2})$, where $\ket{i}$ and $\ket{2,i}$ span the error subspace $\mathcal{H}_{E_i}$. However, these error subspaces are not orthogonal to the code space, reflecting the fact that bit-flip errors cannot be perfectly distinguished from the DM signal~\cite{Chen:2025tgj}. We also note that error subspaces are mutually non-orthogonal, i.e., $\braket{2, i}{2, j} \ne 0$ for $i \ne j$.

Thus, instead of projective syndrome measurements, we employ a positive operator-valued measure (POVM) with the following measurement operators:
\begin{align}
  M_0 &\equiv \ketbra{0}{0} + \ketbra{W}{W}, \\
  M_i &\equiv \sqrt{\frac{N-1}{N}}
  \left( \ket{i}_\perp \!\! \bra{i}_\perp + \ket{2, i}_\perp \!\! \bra{2, i}_\perp\right), \\
  M_S &\equiv \ketbra{S}{S},
\end{align}
for $i=1,2,\ldots,N$, where
\begin{align}
  \ket{i}_\perp &\equiv \sqrt{\frac{N}{N-1}} \ket{i} - \frac{1}{\sqrt{N-1}} \ket{W}, \\
  \ket{2, i}_\perp &\equiv \sqrt{\frac{N}{N-2}} \ket{2, i} - \sqrt{\frac{2}{N-2}} \ket{S}.  
\end{align}
Notice that $|\!\ket{i}_\perp\!|=|\!\ket{2, i}_\perp\!|=1$, and $\braket{W}{i}_\perp=\braket{S}{2, i}_\perp=0$. The corresponding POVM elements are $E_a = M_a^\dagger M_a$ for $a = 0, 1, \ldots, N, S$, and $\tilde{E} = I - \sum_{a} E_a$. As shown in App.~\ref{app:POVM_proof}, these operators form a POVM, and $\text{Tr}[\tilde{E} \rho(t + \Delta t)] = 0$ up to $\mathcal{O}(\Delta t^2)$. Thus, the measurement outcomes are restricted to $\{0, 1, \ldots, N, S\}$.
The measurement outcome $i = 1, 2, \ldots, N$ corresponds to detecting a bit-flip error on the $i$-th sensor. The additional outcome $S$ is required for the completeness.

We next define the recovery operation based on the measurement outcomes.
Errors are detected when the measurement outcome is $a = 1, 2, \ldots, N, S$.
When the outcome is $a = 1, 2, \ldots, N$, we apply such a unitary operation $U_i$ that maps $\ket{i}_\perp$ to $\ket{0}$ and $\ket{2, i}_\perp$ to $\ket{W}$, which approximately corrects the bit-flip error on the $i$-th sensor.
When the outcome is $S$, this is originated from a bit-flip error on $\ket{W}$, and we apply a unitary operation $U_S$ that maps $\ket{S}$ to $\ket{W}$.

Combining the syndrome measurement and the recovery operation,
{we can write the whole control operation as new measurement operators; i.e.,} $R_0 \equiv M_0$ and
\begin{align}
  R_i &\equiv U_i M_i = \sqrt{\frac{N-1}{N}} \left( \ketbra{0}{i}_\perp + \ketbra{W}{2, i}_{\perp} \right)\!, \\
  R_S &\equiv U_S M_S = \ketbra{W}{S}.
\end{align}
Note that $R_a^\dagger R_a = E_a$ for all $a$. The control operation $\mathcal{E}_{\rm c}$ is then defined as
\begin{align}
  \mathcal{E}_{\rm c}[\rho] \equiv \sum_{a} R_a \rho R_a^\dagger.
  \label{eq:control_operation}
\end{align}
Obviously, states after the control operation lie in $\mathcal{H}_L$.
Throughout this control operation, we need to use $\mathcal{O}(\log N)$ ancilla qubits for implementing the POVM\,\cite{stinespring1955positive} as projection measurements. This is small enough to be negligible when we discuss the scaling.

Let us now see the density matrix after a single unit interval corresponding to time $t$ to $t + \Delta t$. Using Eqs.\,(\ref{eq:channel_E_dt}) and (\ref{eq:control_operation}), we have
\begin{align}
  \mathcal{E}_{\rm c} \circ \mathcal{E}_{\Delta t}^{\otimes N}[\rho(t)] &\simeq \rho(t) + \mathcal{E}_{\rm c}\qty[\mathcal{L}[\rho(t)]] \Delta t,
\end{align}
up to $\mathcal{O}(\Delta t^2)$. Then,
\begin{align}
  \mathcal{E}_{\rm c}\qty[\mathcal{L}[\rho(t)]] &= -i [H_{IL}, \rho(t)] \nonumber \\ 
  & \quad - \frac{N \gamma}{2} \rho(t) + \frac{\gamma}{2} \sum_{a} \sum_{i=1}^N \mathscr{S}_{a,i}^x \rho(t) \mathscr{S}_{a,i}^{x \dagger},
\end{align}
where
\begin{align}
  H_{IL} &\equiv R_0 H_I R_0^\dagger = \epsilon \sqrt{N} \qty( \sigma_L^x \cos \varphi  - \sigma_L^y \sin \varphi ), \\ 
  \mathscr{S}_{a,i}^x &\equiv R_a \sigma_i^x R_0.
\end{align}
Here, $\sigma_L^a$ ($a = x, y, z$) are the Pauli operators acting on the logical code space.
A straightforward calculation yields
\begin{widetext}
\begin{align}
  \mathcal{E}_{\rm c}\qty[\mathcal{L}[\rho(t)]] &= -i [H_{IL}, \rho(t)] + \frac{\gamma}{2} \qty[-\rho(t) + \sigma_L^x \rho(t) \sigma_L^x] + \frac{c \gamma}{4} \qty[ -\rho(t) + \sigma_L^z \rho(t) \sigma_L^z] \equiv \mathcal{L}_L[\rho(t)],
  \label{eq:effective_Lindblad}
\end{align}
\end{widetext}
where we have defined $c \equiv \frac{2(N - 1)}{N + \sqrt{N(N - 2)}} = \mathcal{O}(N^0)$, and
\begin{align}
  \mathcal{E}_{\rm c} \circ \mathcal{E}_{\Delta t}^{\otimes N}[\rho(t)] &\simeq \rho(t) + \mathcal{L}_L[\rho(t)] \Delta t,
\end{align}
up to $\mathcal{O}(\Delta t^2)$.
$\mathcal{L}_L$ describes the effective time evolution of the density matrix within the code space and can be interpreted as an effective Lindblad superoperator for the logical qubit.

\textit{Uncertainty Estimation ---}
\label{sec:uncertainty}
We now estimate the uncertainty of the DM signal with our QEC-like protocol.
Let the total time we use be $T$, during which we repeat multiple measurements with time $\tau$.
For the moment, we assume 
that the phase $\varphi$ can be treated as a constant during $T$.
As we have seen, after a single unit interval, the density matrix evolves in the code space according to the effective Lindblad equation with $\mathcal{L}_L$ up to $\mathcal{O}(\Delta t^2)$.
As we take the limit ${N\gamma}\Delta t \to 0$, the density matrix at time $T$ is arbitrarily close to the solution of the effective Lindblad equation
\begin{align}
  \frac{d\rho_L}{dt} = \mathcal{L}_L[\rho_L],
\end{align}
with the initial condition $\rho_L(0) = \ket{0}_L\!\!\bra{0}_L$. To estimate the DM signal $\epsilon$, we perform a projective measurement $P_W \equiv \ket{1}_L\!\!\bra{1}_L$ (or, for the states in the logical code space, equivalently, $\sum_i \ket{1}_i\!\!\bra{1}_i$ on the physical qubits) 
on the final state $\rho_L(T)$.

To discuss the performance of measurement protocols, we use the standard deviation of the estimator of $\epsilon$ as a measure of uncertainty.
Using the linear approximation\,\cite{Fukuda:2025zcf}, the standard deviation, $\delta \epsilon$, is given by
\begin{align}
  \delta \epsilon \simeq \frac{\sqrt{\ev{O^2} - \ev{O}^2}}{\abs{\pdv{\ev{O}}{\epsilon}} \sqrt{N_\text{rep}}},
  \label{eq:uncertainty_general}
\end{align}
where $O$ is the observable to estimate $\epsilon$ and $N_\text{rep}$ is the number of repetitions of measurements, given as $T / \tau$.
The expectation values are calculated with respect to the final state after each sensing time $\tau$, which is determined so that $\delta \epsilon$ is minimized for each protocol.

{
First, let us estimate the uncertainty without our QEC-like protocol to see how the excitation noise degrades the sensitivity.
Starting from the initial state $\rho_0(0) = \ket{0}\!\!\bra{0}$, the uncertainty, $\delta \epsilon_0$, is calculated using the density matrix $\rho_0(t)$, which is the solution of the original Lindblad equation, Eq.\,(\ref{eq:lindblad}).
The minimal uncertainty is predicted by the QCRB; as done in App.\,\ref{app:QFI}, we find
\begin{align}
  \label{eq:QCRB_noQEC}
  \delta \epsilon_0 \gtrsim \sqrt{\frac{\gamma}{N T}}.
\end{align}
The QCRB is indeed saturated by the Pauli-$Y$ measurements on each sensor, $O = P_Y \equiv \sum_{i=1}^N \frac12 (1 + \sigma_i^y)$. $\ev{P_Y}_{\rho_0(\tau)} \simeq \frac{N}{2} \qty(1 - \epsilon \tau \cos \varphi)$ and
the uncertainty is indeed $\delta \epsilon_0^{(Y)} \sim \sqrt{\gamma / N T}$ for $\tau \sim 1/\gamma$.
However, the DM phase is unknown and can be treated as a random variable uniformly distributed in $[0, 2\pi)$, which changes over time scales longer than the DM coherence time.
If we average over $\varphi$, the expectation value of $P_Y$ becomes independent of $\epsilon$, making it impossible to estimate the DM signal.
Therefore, we cannot repeat the ideal measurement over long time scales in practice.
}

{In this situation, the best we can do without our QEC-like protocol is rather to}
count the total number of excited sensors, i.e., $O = P_Z \equiv \sum_{i=1}^N \frac12 (1 - \sigma_i^z)$\,\cite{Chen:2022quj}.
As a result, the expectation value is given by
\begin{align}
  \ev{P_Z}_{\rho_0(\tau)} &\simeq N \qty(c_0^{(S)} \epsilon^2 \tau^2 + c_0^{(N)} \gamma \tau),
  \label{eq:expectation_noQEC}
\end{align}
up to the leading order in $\epsilon^2 \tau^2$ and $\gamma \tau$, where $c_0^{(S)}$ and $c_0^{(N)}$ are constants of order unity.
Using Eq.\,(\ref{eq:uncertainty_general}), we obtain
\begin{align}
  \delta \epsilon_0 \sim \frac{\sqrt{\gamma}}{\epsilon \tau \sqrt{N T}} \sim \frac{\gamma^{3/2}}{\epsilon \sqrt{N T}},
  \label{eq:uncertainty_noQEC}
\end{align}
where we have chosen $\tau \sim 1/\gamma$ to minimize $\delta \epsilon_0$.
{This uncertainty is larger than the QCRB, Eq.\,\eqref{eq:QCRB_noQEC}, by a factor of $\gamma / \epsilon$, which can be significant as we are interested in the regime $\gamma \gg \epsilon$.
The reason why Pauli-$Y$ measurements are better than Pauli-$Z$ measurements is as follows.
Generally, Pauli-$Y$ measurements give larger signal effects, $\mathcal{O}(\epsilon \tau)$, compared to Pauli-$Z$ measurements, $\mathcal{O}(\epsilon^2 \tau^2)$, but without noise, this is compensated by the larger variance, resulting in the same uncertainty scaling. However, with the bit-flip noise, both measurements result in the larger variance due to the noise, but since the signal effect is larger for Pauli-$Y$ measurements, the uncertainty scaling is better.}

We next estimate the uncertainty with our QEC-like protocol, denoted as $\delta \epsilon_{L}$ with $O = P_W$. The effective Lindblad equation contains both bit-flip and dephasing noises, limiting the sensing time to $\tau \lesssim 1/\gamma$ as in the previous case.
To estimate $\delta \epsilon_{L}$, we need to compare the signal rate, $\epsilon \sqrt{N}$, with the noise rate, $\mathcal{O}(N^0) \times \gamma$.
With fixed $\epsilon$ and $\gamma$ with $\epsilon \ll \gamma$, the signal rate is much smaller than the noise rate for small $N$, $N \lesssim N_\text{th} \equiv (\gamma / \epsilon)^2$.
Then, the expectation value is given by
\begin{align}
  \ev{P_W}_{\rho_L(\tau)}^\text{(small $N$)} &\simeq c_L^{(S)} \epsilon^2 N \tau^2 + c_L^{(N)} \gamma \tau,
\end{align}
up to the leading order in $\epsilon^2 N \tau^2$ and $\gamma \tau$, where $c_L^{(S)}$ and $c_L^{(N)}$ are constants of order unity.
Using Eq.\,(\ref{eq:uncertainty_general}), we obtain
\begin{align}
  \delta \epsilon_{L}^\text{(small $N$)} \sim \frac{\gamma^{3/2}}{\epsilon N \sqrt{T}},
  \label{eq:uncertainty_QEC_smallN}
\end{align}
for small $N$, where we have chosen $\tau \sim 1/\gamma$ to minimize the uncertainty.
Comparing this with Eq.\,(\ref{eq:uncertainty_noQEC}), we find that our protocol improves the uncertainty by a factor of $\sqrt{N}$ for small $N$. 

For $N \gtrsim N_\text{th}$, the signal rate becomes larger than the noise rate. In this case, we may ignore the noise contribution, yielding
\begin{align}
  \ev{P_W}_{\rho_L(\tau)}^\text{(large $N$)} &\simeq \sin^2 \epsilon \sqrt{N} \tau \sim \epsilon^2 N \tau^2,
\end{align}
The uncertainty is then given by
\begin{align}
  \delta \epsilon_{L}^{\prime \text{(large $N$)}} \sim \frac{1}{\sqrt{N \tau T}} \sim \frac{\sqrt{\epsilon}}{N^{1/4} \sqrt{T}},
  \label{eq:uncertainty_QEC_largeNPrime}
\end{align}
where we have chosen $\tau \sim 1/\epsilon \sqrt{N}$. However, this is not optimal; rather, if we divide $N$ sensors into $N / N_\text{th}$ groups of size $N_\text{th}$ and apply our protocol to each group independently, the number of repetitions increases by $N / N_\text{th}$ and we obtain
\begin{align}
  \delta \epsilon_{L}^\text{(large $N$)} \sim \frac{1}{\sqrt{N / N_\text{th}} \cdot \sqrt{N_\text{th} \tau T}} \sim \sqrt{\frac{\gamma}{NT}},
  \label{eq:uncertainty_QEC_largeN}
\end{align}
where we have chosen $\tau \sim 1/\gamma$. One can check $\delta \epsilon_{L}^\text{(large $N$)} < \delta \epsilon_{L}^{\prime \text{(large $N$)}}$ for $N > N_\text{th}$.
{Therefore, our protocol achieves the same uncertainty as the ideal measurement without error mitigation for large $N$, which is unachievable without error mitigation as the DM phase is unknown.}

\begin{figure}[tbp]
  \centering
  \hspace{-0.5cm}\includegraphics[width=0.46\textwidth]{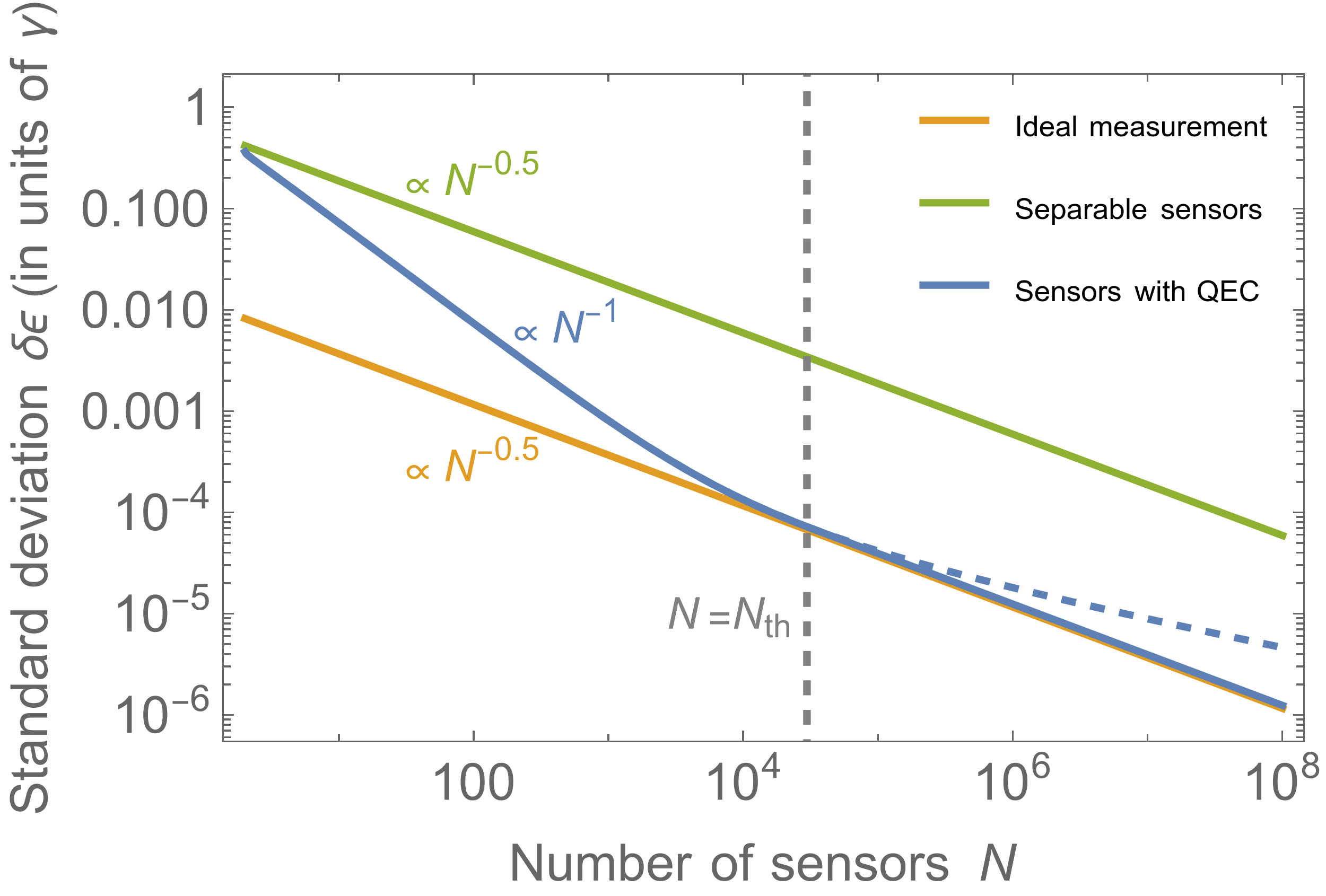}
  \caption{Numerical calculation of the uncertainty with our QEC-like protocol. The blue {dashed} line shows the uncertainty with our protocol, calculated by numerically solving the effective Lindblad equation, Eq.\,(\ref{eq:effective_Lindblad}). The blue {solid} line adopts the optimal strategy for large $N$ by dividing sensors into groups of size $N_\text{th}$. 
  The green and orange lines show the numerical uncertainty of measurements without error mitigation with $O = P_Z$ and $O = P_Y$, respectively. {The gray vertical dashed line shows the numerically optimized threshold $N = N_\text{th}$.}}
  \label{fig:numerical}
\end{figure}

In Fig.\,\ref{fig:numerical}, we show the results of the numerical calculation of the uncertainty with our QEC-like protocol and compare it with the uncertainty of measurements without error mitigation and Pauli-$Y$ measurements. Here, we take $\varphi = 0, \epsilon = 10^{-2} \gamma$ and $T = {10^4}/\gamma$ and numerically optimize the sensing time $\tau$ to minimize the uncertainty for each $N$. Indeed, we can see that the uncertainty scales with $1/N$ for small $N$.
For large $N$, as we see in Eqs.\,\eqref{eq:uncertainty_QEC_largeNPrime} and \eqref{eq:uncertainty_QEC_largeN}, the uncertainty scaling changes around $N_\text{th} \simeq (\gamma / \epsilon)^2$.
We numerically choose $N_\text{th} = \mathcal{O}((\gamma / \epsilon)^2)$ such that the uncertainty is minimized if we use the optimal strategy for $N > N_\text{th}$.
We see that our protocol achieves the same scaling as the measurement with $O = P_Y$.

Lastly, we comment on the case where we have other types of noise than the bit-flip noise. Suppose, for simplicity, we have both excitation and de-excitation noises with the rate $\gamma_\text{EX}$ and $\gamma_\text{DE}$, respectively. As explained in App.\,\ref{app:reduction_to_bit_flip}, we can reduce these noises to the bit-flip noise with rate $\gamma_\text{eff} \sim \gamma_\text{EX} + \gamma_\text{DE}$ by constructing logical qubits from 3 physical qubits and applying control operations. Assuming $\epsilon \ll \gamma_\text{EX}$, the uncertainty without error mitigation is then given as\,\cite{Chen:2025tgj}
\begin{align}
  \delta \epsilon_0^{(\text{EX$+$DE})} \sim \frac{\sqrt{\gamma_\text{EX}} \gamma_\text{eff}}{\epsilon \sqrt{N T}}.
\end{align}
On the other hand, with our protocol, the uncertainty is given by Eqs.\,(\ref{eq:uncertainty_QEC_smallN}) and (\ref{eq:uncertainty_QEC_largeN}) with $\gamma \to \gamma_\text{eff}$. Therefore, our protocol improves the uncertainty for this case as well. 

\textit{Conclusions and Discussion ---}
\label{sec:conclusion}
In this letter, {we have addressed a key challenge in quantum-enhanced 
searches for wave-like dark matter: the presence of an unknown, 
time-varying signal phase that prevents the repeated application of 
ideal measurements over long observation times.}
We proposed a QEC-like error mitigation protocol to improve the sensitivity of quantum sensors for DM searches under general Markovian noise acting on each sensor independently. By constructing a logical qubit from $N$ sensors and applying adaptive control operations, our protocol improves the uncertainty by a factor of $\sqrt{N}$ compared to the protocol without error mitigation for small $N$. For large $N$, the uncertainty scaling saturates to the SQL by the ideal measurement, {without requiring 
knowledge of the signal phase.}
{To our knowledge, this is the first 
{proposal} that the ideal-measurement sensitivity can be recovered 
in the practically relevant regime of unknown and drifting phase, a 
scenario ubiquitous in DM searches or other high-energy physics applications, such as high-frequency gravitational wave detection,
 but often overlooked in conventional 
quantum metrology.}


In practice, we admit that our protocol is complicated and challenging to implement with current technology {for large $N$}. It is an important future work to simplify the protocol {possibly by using hardware-efficient recovery operations} and to analyze the performance under imperfect quantum operations and finite operation time for practical applications.


\textit{Acknowledgments ---}
The work of TS was supported by the JSPS fellowship Grant No.\ 23KJ0678.
This work was supported by JSPS KAKENHI Grant Nos.\ 24K17042 [HF], 25H00638 [HF], 26H00403 [HF], 26K17131 [HF] and 23K22486 [TM].
In this research work, HF and TM used the UTokyo Azure (\url{https://utelecon.adm.u-tokyo.ac.jp/en/research_computing/utokyo_azure/}).

\appendix
\section{Notation}
\label{app:notation}
In this appendix, we summarize our notation for multi-qubit states used in this letter.
The ground state of $N$ qubits is written as
\begin{align}
  \ket{0} \equiv \bigotimes_{i=1}^N \ket{0}_i
\end{align}
Single-excitation states are labeled by the position of the excited qubit:
\begin{align}
  \ket{i} \equiv \sigma_i^x \ket{0},
\end{align}
where $i=1,2,\ldots,N$.
The $W$ state is given by
\begin{align}
  \ket{W} \equiv \frac{1}{\sqrt{N}} \sum_{i=1}^N \ket{i}.
\end{align}
Double-excitation states, with excitations at qubits $i$ and $j$, are expressed as
\begin{align}
  \ket{i,j} \equiv \sigma_i^x \sigma_j^x \ket{0},
\end{align}
for $i \ne j$ and $i,j=1,2,\ldots,N$.
The equal superposition of double-excitation states with one excitation fixed at qubit $i$ is
\begin{align}
  \ket{2, i} \equiv \frac{1}{\sqrt{N - 1}} \sum_{\substack{j=1 \\ j \neq i}}^N \ket{i,j},
\end{align}
for $i=1,2,\ldots,N$.
The symmetric superposition of all double-excitation states is
\begin{align}
  \ket{S} \equiv \frac{\sqrt{2}}{\sqrt{N (N-1)}} \sum_{1 \leq i < j \leq N} \ket{i,j} = \frac{1}{\sqrt{2 N}} \sum_{i=1}^N \ket{2, i}.
\end{align}
All states above are normalized.

\section{Reduction of Various Noises to Bit-Flip Noise}
\label{app:reduction_to_bit_flip}
In this appendix, we explain how various types of noise can be reduced to the bit-flip noise considered in the main text.
As we have mentioned, we first construct a logical qubit by using 3 physical qubits to form the GHZ state and then apply QEC operations.
We divide the total sensing time into small intervals, $\Delta t_3$, let the system evolve under the noise for a short time, and then apply a control operation.

The logical code space is defined as 
\begin{align}
  \mathcal{H}_{3L} = \{\ket{+}_{3L} \equiv \ket{+++},\, \ket{-}_{3L} \equiv \ket{---}\},
\end{align}
where $\ket{\pm} = (\ket{0} \pm \ket{1})/\sqrt{2}$. The initial state is $\ket{\psi_{3, \rm init}} = \ket{0}_{3L}$, where 
\begin{align}
  \ket{0}_{3L} &\equiv \frac{1}{\sqrt{2}} \qty(\ket{+}_{3L} + \ket{-}_{3L}) \\ 
  \ket{1}_{3L} &\equiv \frac{1}{\sqrt{2}} \qty(\ket{+}_{3L} - \ket{-}_{3L}).
\end{align}
For short enough $\Delta t_3$, we may assume that at most one noise event occurs during each interval. Therefore, depending on which sensor the noise acts on, three types of error subspaces appear:
\begin{align}
  \mathcal{H}_{3E, i} = \{\sigma_i^z \ket{+}_{3L},\, \sigma_i^z \ket{-}_{3L}\},
\end{align}
for $i=1,2,3$. Let $P_i$ be the projection operator onto $\mathcal{H}_{3E, i}$ and $P_L$ be the projection operator onto the code space $\mathcal{H}_{3L}$. To detect errors, we perform the projective measurement with the measurement operators $\{P_L, P_1, P_2, P_3\}$. If the measurement outcome is $P_i$, we apply the recovery operation $\sigma_i^z$ to correct the error. Let us denote this control operation as $\mathcal{E}_{3\rm c}$.

Now, we consider the effect of various types of noise, as well as the DM signal, during the time interval $\Delta t_3$.
For simplicity, we focus on de-excitation noise, excitation noise, and the dephasing noise on individual sensors, but other types of noise can be treated similarly.
Similar to the main text, we start from the density matrix $\rho_{3}(t)$ with the three qubits at time $t$ within the code space and let the system evolve under the total Lindblad equation for a time duration $\Delta t_3$. Then, the density matrix at time $t + \Delta t_3$ is given by
\begin{align}
  \rho_{3}(t + \Delta t_3) &\simeq \rho_{3}(t) + \mathcal{L}_3[\rho_{3}(t)] \Delta t_3,
\end{align}
up to first order in $\Delta t_3$. The Lindblad superoperator $\mathcal{L}_3$ is defined as
\begin{align}
  \mathcal{L}_3[\rho_3] &\equiv -i [H_{3I}, \rho_3] + \sum_{i=1}^3 \sum_{\alpha} \gamma_\alpha D_i^{(\alpha)}[\rho_3],
\end{align}
where $H_{3I}$ is the sum of the interaction Hamiltonian in the interaction picture for the three qubits and
the sum over $\alpha$ runs over DE (de-excitation), EX (excitation), and DP (dephasing)
with $\gamma_{\rm DE}$, $\gamma_{\rm EX}$, and $\gamma_{\rm DP}$ being these noise rates, respectively.
The superoperators $D_i^{(\text{DE})}$, $D_i^{(\text{EX})}$, and $D_i^{(\text{DP})}$ are defined as
\begin{align}
  D_i^{(\text{DE})}[\rho_3] &\equiv \frac{1}{2} \qty( a_i \rho_3 a_i^\dagger - \frac{1}{2}\{a_i^\dagger a_i,\rho_3 \}), \\
  D_i^{(\text{EX})}[\rho_3] &\equiv \frac{1}{2} \qty( a_i^\dagger \rho_3 a_i - \frac{1}{2}\{a_i a_i^\dagger,\rho_3 \}), \\
  D_i^{(\text{DP})}[\rho_3] &\equiv \frac{1}{2} \qty( \sigma_i^z \rho_3 \sigma_i^z - \rho_3),
\end{align}
where $a_i \equiv \ket{0}_i\!\! \bra{1}_i$ is the annihilation operator for the $i$-th sensor.

After applying the control operation $\mathcal{E}_{3\rm c}$, 
\begin{align}
  \mathcal{E}_{3\rm c}\qty[\mathcal{L}_3[\rho_{3}(t)]] &= -i [H_{3LI}, \rho_3(t)] \nonumber \\ 
  & \quad+ \sum_{i}^3 \sum_{\alpha} \gamma_\alpha \mathcal{E}_{3\rm c}\qty[D_i^{(\alpha)}[\rho_{3}(t)]],
\end{align}
where
\begin{align}
  H_{3LI} &\equiv P_L H_I P_L = 3\epsilon \sigma_{3L}^x \cos\varphi,
\end{align}
with $\sigma_{3L}^a$ ($a = x, y, z$) being the Pauli operators acting on the logical code space, $\{\ket{0}_{3L}, \ket{1}_{3L}\}$.
Although the oscillation term proportional to $\sigma_{3L}^y$ vanishes, the effective coupling strength is of the same order as the original one.
By a straightforward calculation, we obtain
\begin{align}
  \mathcal{E}_{3, \rm c}\qty[D_i^{(\text{DE})}[\rho_{3}(t)]] &= \frac{1}{4} \qty[-\rho_{3}(t) + \sigma_{3L}^x \rho_{3}(t) \sigma_{3L}^x], \\
  \mathcal{E}_{3, \rm c}\qty[D_i^{(\text{EX})}[\rho_{3}(t)]] &= \frac{1}{4} \qty[-\rho_{3}(t) + \sigma_{3L}^x \rho_{3}(t) \sigma_{3L}^x], \\
  \mathcal{E}_{3, \rm c}\qty[D_i^{(\text{DP})}[\rho_{3}(t)]] &= 0.
\end{align}
Therefore, the effective noise after the control operation is described by the bit-flip noise with the rate $3(\gamma_{\rm DE} + \gamma_{\rm EX})/2$. These procedures may change the signal strength, the noise rate and the number of sensors by factors of order unity, but the scaling of the uncertainty remains unchanged.

\section{Positivity of Our POVM Elements}
\label{app:POVM_proof}
In this appendix, we prove that our POVM elements, $E_a$ ($a = 0, 1, \ldots, N, S$), and $\tilde{E} = I - \sum_a E_a$ are positive operators. It is obvious that $E_a$ are positive operators since they are defined as $E_a = M_a^\dagger M_a$. Therefore, we only need to show that $\tilde{E}$ is a positive operator.

To show this, we first separate the Hilbert space by the total excitation numbers.
Let $P_k$ be the projection operator onto the subspace with $k$ excitations.
By the definition, we have $E_a P_k = P_k E_a = 0$ for $k > 2$ and any $a$. Therefore, we need to prove that $P_k \tilde{E} P_k$ are positive semidefinite for $k = 0, 1, 2$.
For $k = 0$, we have $P_0 \tilde{E} P_0 = P_0 - \ketbra{0} = 0$. For $k = 1$, 
\begin{align}
  P_1 \tilde{E} P_1 = P_1 - \ketbra{W} - \frac{N-1}{N} \sum_{i=1}^N \ket{i}_\perp \!\! \bra{i}_\perp = 0,
\end{align}
where we have used $P_1 = \sum_{i=1}^N \ketbra{i}$.

Finally, we consider the $k = 2$ case. Using the definition of $E_a$, we obtain
\begin{align}
  P_2 \tilde{E} P_2 &= P_2 - \frac{N - 1}{N - 2} \sum_{i=1}^N \ketbra{2, i} + \frac{N}{N - 2} \ketbra{S}.
\end{align}
Let a matrix $G$ be defined as $G \equiv \sum_{i=1}^N \ketbra{2, i}$. The kernel of $G$ in the double-excitation subspace is an orthogonal complement of the space $\mathcal{H}_G \equiv \{\ket{2, i}\}_{i=1}^N$. Since $\ket{S}$ is in $\mathcal{H}_G$, we need to focus on $\mathcal{H}_G$ to check the positivity of $P_2 \tilde{E} P_2$. Also, it is easy to see that $\ket{S}$ is an eigenvector of $G$ with the eigenvalue $2$.

To see the eigenvalues of $G$, we assume $\ket{2, i}$ as a matrix with the row index being the index of the Hilbert space and the column index being $i$. Let us denote this matrix as $\mathcal{A}$. Then, $G$ can be written as $G = \mathcal{A} \mathcal{A}^\dagger$. By the singular value decomposition, we can write $\mathcal{A} = U \Sigma V^\dagger$, where $U$ and $V$ are unitary matrices and $\Sigma$ is a diagonal matrix with non-negative real numbers on the diagonal. Therefore, the non-zero eigenvalues of $G$ are given by the square of the non-zero singular values of $\mathcal{A}$, which are equal to the non-zero eigenvalues of $\tilde{G} \equiv \mathcal{A}^\dagger \mathcal{A}$. The $i,j$ component of $\tilde{G}$ is given by $\braket{2, i}{2, j}$ and therefore
\begin{align}
  \tilde{G} = \frac{N - 2}{N - 1} I + \frac{1}{N - 1} J,
\end{align}
where $J$ is the matrix with all components being $1$. The eigenvalues of $J$ are $0$ with the multiplicity of $N - 1$ and $N$ with the multiplicity of $1$. Therefore, the eigenvalues of $\tilde{G}$ are $2$ and $\frac{N - 2}{N - 1}$ with the multiplicity of $1$ and $N - 1$, respectively. As a result, so are the eigenvalues of $G$. Since $\ket{S}$ is the eigenvector with the eigenvalue $2$, we obtain 
\begin{align}
  P_2 \tilde{E} P_2 \big|_{\mathcal{H}_G} = 0,
\end{align}
which completes the proof. It is straightforward to see that $P_2\mathcal{E}_{\Delta t}^{\otimes N}[\rho(t)]P_2$ is an operator in $\mathcal{H}_G$ for any density matrix $\rho(t)$ in the code space. Therefore, we may also see that $\tilde{E} \mathcal{E}_{\Delta t}^{\otimes N}[\rho(t)]$ is vanishing.


\section{Quantum Fisher Information of the Original Lindblad Equation}
\label{app:QFI}
In this appendix, we calculate the quantum Fisher information (QFI) of the original Lindblad equation, Eq.\,(\ref{eq:lindblad}), with the initial state $\ket{0}$.

Similar to the main text, we assume the measurement time $\tau$ is short enough so that we can expand the density matrix up to the first order in $\epsilon \tau$ and $\gamma \tau$.
The density matrix at time $\tau$ is given by
\begin{align}
  \rho_0(\tau) &= \bigotimes_{i=1}^N \rho_i(\tau), \\ 
  \rho_i(\tau) &\simeq \ket{0}_i\!\!\bra{0}_i - \epsilon \tau \qty(\sigma_i^y \cos \varphi + \sigma_i^x \sin \varphi) - \frac{\gamma \tau}{2} \sigma_i^z.
\end{align}
The eigenvectors of $\rho_i(\tau)$ are given by $\ket{0}_i$ and $\ket{1}_i$ with the eigenvalues $1$ and $0$, respectively, up to $\mathcal{O}(\epsilon \tau, \gamma \tau)$.
Using these expressions, we can calculate the QFI for a single sensor, $\mathcal{F}_i$, as
\begin{align}
  \mathcal{F}_i \sim \abs{\!\mel{0}{\partial_\epsilon \rho_i(\tau)}{1}}^2  \sim \tau^2.
\end{align}
Therefore, the total QFI for $N$ sensors and the total measurement time $T$ is given by
\begin{align}
  \mathcal{F}_\text{total} \sim N \frac{T}{\tau} \tau^2 = N \tau T.
\end{align}
Using the QCRB, the standard deviation of the estimator of $\epsilon$ is bounded as
\begin{align}
  \delta \epsilon \geq \frac{1}{\sqrt{\mathcal{F}_\text{total}}} \sim \frac{1}{\sqrt{N \tau T}} \sim \sqrt{\frac{\gamma}{N T}},
\end{align}
where we have chosen $\tau \sim 1/\gamma$.

\bibliography{papers}


\end{document}